# Fluorescence-Detected Mid-Infrared Photothermal Microscopy

Minghe Li[1†], Aleksandr Razumtcev[1†], Ruochen Yang[2], Youlin Liu[1], Jiayue Rong[1], Andreas C. Geiger[1], Romain Blanchard[3], Cristian Pflueg[3], Lynne S. Taylor[2], and Garth J. Simpson[1*].

[1]Department of Chemistry, Purdue University, West Lafayette IN 47907. [2]Physical and Industrial Pharmacy, Purdue University, West Lafayette IN 47907. [3]Pendar Technologies, 30 Spinelli Pl, Cambridge MA 02138.

†Contributed equally.

*Supporting Information*

**ABSTRACT:** We demonstrate instrumentation and methods to enable fluorescence-detected photothermal infrared (F-PTIR) microscopy, then demonstrate the utility of F-PTIR to characterize the composition within phase-separated domains of model amorphous solid dispersions (ASDs) induced by water sorption. In F-PTIR, temperature-dependent changes in fluorescence quantum efficiency are shown to sensitively report on highly localized absorption of mid-infrared radiation. The spatial resolution with which infrared spectroscopy can be performed is dictated by fluorescence microscopy, rather than the infrared wavelength. Following proof of concept F-PTIR demonstration on model systems of polyethylene glycol (PEG) and silica gel, F-PTIR enabled the characterization of chemical composition within inhomogeneous ritonavir / polyvinylpyrrolidone-vinyl acetate (PVPVA) amorphous dispersions. Phase separation is implicated in the observation of critical behaviors in ASD dissolution kinetics, with the results of F-PTIR supporting the formation of phase-separated drug-rich domains upon water absorption in spin-cast films.

Photothermal infrared microscopy has recently emerged as a powerful tool for high-resolution optical imaging, with numerous advantages relative to alternative methods for pharmaceutical materials analysis. Photothermal microscopy with a spatial resolution on the order of a few nm has been realized by photothermal atomic force microscopy infrared spectroscopy (AFM-IR),[1-2] in which heat-induced perturbation to an atomic force microscope tip informs on surface absorption. Photothermal AFM-IR is now a mature technology, routinely capable of providing mid-infrared microspectroscopy with nm-scale spatial resolution,[3] with growing use in analysis of pharmaceutically relevant materials.[4] However, it suffers from two key limitations; i) like most scanning probe microscopy methods, the mechanical response time of the cantilever sets a practical speed limit on pixel and frame rates, often requiring several minutes for a single image, and ii) by design, it intrinsically can only inform on absorption in the region immediately adjacent to the surface.

Several of the limitations of photothermal AFM-IR can be overcome by using light to transduce the photothermal temperature change. In optically detected photothermal infrared (O-PTIR), subtle changes in refractive index induced by local temperature changes from absorption of an IR beam are recorded through perturbation of a co- or counter-propagating visible beam.[5] For pharmaceutical materials analysis, O-PTIR has been used by Li *et al.* in an epi-detection configuration compatible with analysis of powders and opaque solid-state materials.[6] In brief, a mid-infrared beam from a quantum cascade laser (QCL) was co-propagated with a 785 nm visible beam using a reflective Cassegrain objective. IR-induced modulation in the back-scattered visible light was detected using a polarizing beam-splitting cube, exploiting depolarization by the highly turbid solid-state samples. IR-modulated changes in the backscattering enabled spectral assignment of active and inactive ingredients in a Tylenol tablet.

Despite these successes, O-PTIR measured by perturbation of a probe beam suffers from several limitations, exacerbated in analysis of pharmaceutical materials. First, refractive index is a fairly weak function of temperature, changing by ~0.01% per °C for water at room temperature.[7] Second, it is well known that the photothermally induced deflection of a probe beam in transmission approaches zero for a point source centered in the probe beam focus and is maximized with an axial offset between the probe beam and photothermal lens of $\sqrt{3}Z_0$ ( $Z_0$ is the Rayleigh range, equal to half the depth of focus).[8-9] As such, an O-PTIR measurement of a point source along the optical axis based on detection of beam deflection is only expected to produce peak signals in homogeneous media when the probe beam focus is displaced by nearly a full depth of field relative to the photothermal source, potentially producing nontrivial 3D point spread functions.

For measurements of transparent samples acquired in transmission, this limitation can be overcome by coupling O-PTIR with quantitative phase contrast approaches, such as those developed by Popescu and coworkers[10-12], enabling photothermal contrast through interferometry rather than beam deflection.[13] Dual-path interferometric approaches have also been shown by Orrit and coworkers to enable sensitive photothermal visible-wavelength absorption spectroscopy of particles in solution, approaching single molecule sensitivity.[14] However, many pharmaceutical materials exhibit significant turbidity and heterogeneity, creating challenges for analytically modeling and interpreting photothermal changes in back-scattered intensity from chemically and physically heterogeneous samples. In many instances, the optical constants associated with temperature change are not known, complicating quantitative interpretation of the magnitude of O-PTIR responses. Furthermore, the detected back-scattered signal can depend on both depolarization and scattering in ways that may depend nontrivially on the particle size distribution within compacted or powdered samples.

Alternative optical method for thermal reporting may provide complementary advantages to current strategies based on refractive index detection. In particular, we explore herein the potential viability of using fluorescence intensity to detect local changes in temperature induced by mid-infrared absorption. The quantum efficiency of fluorescence is well established to vary sensitively with temperature.[15] Following optical excitation, increases in temperature enable access to a greater suite of thermally accessible relaxation pathways, all of which compete with fluorescence for a net reduction in fluorescence intensity. Temperature-dependent changes in autofluorescence emission have been reported in microscale thermophoresis, which target measurements of protein mobility and thermal stability.[16-19] In related work, Gruebele and coworkers have used the temperature-dependence of both native autofluorescence[20] and labeled fluorescence[21-22] to recover timescales for protein dynamics in temperature-jump experiments. Compared to refractive index changes, the change in fluorescence quantum efficiency can be quite large, routinely changing by ~2-3%/°C for tryptophan[23] and ~1-2%/°C for aqueous solutions of rhodamine B,[24] corresponding to ~100-fold higher relative change than refractive index detection. Given the high signal to noise with which fluorescence measurements are regularly recorded, this sensitivity has the potential to be more than sufficient to enable fluorescence detection of local temperature perturbations of < 1°C induced by mid-infrared absorption. Following initial proof-of-concept measurement supporting the viability of fluorescence-detected photothermal infrared (F-PTIR) microscopy, we demonstrate its utility in unambiguous observation of water-induced amorphous amorphous phase separation (AAPS) in amorphous solid dispersions of ritonavir in poly-vinyl pyrrolidone/vinyl acetate (PVPVA), with direct implications in pharmaceutical materials design.

The F-PTIR microscope consisted of a continuous 532 nm green laser (SLOC Lasers) co-propagating with a broadly-tunable QCL array (Pendar Technologies) (**Figure 1**).

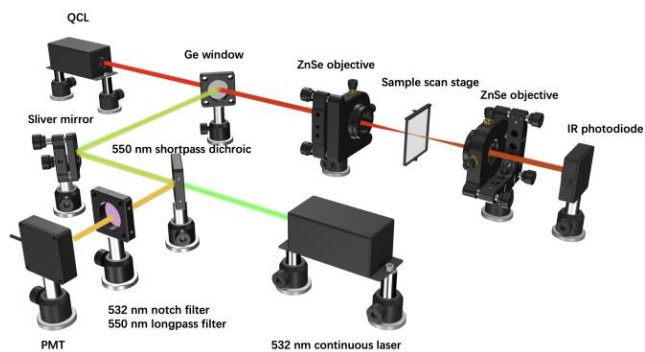

**Figure 1.** F-PTIR microscope configuration, with the mid-infrared QCL beam copropagating with a 532 nm visible beam for fluorescence excitation. The fluorescence signal was collected by a PMT in an epi-detected configuration.

The average power of the green laser was around 0.2 mW on the sample plane for rhodamine-6G (R6G) associated samples and 30 mW for samples associated with Nile red as the fluorescent reporter. The monolithic QCL included 32 independently addressable wavelength channels in the range of wavenumbers from 1190 cm$^{-1}$ to 1330 cm$^{-1}$.[25] In typical operation, the QCL array was operated in a burst mode, in which a rapid series of 300 ns laser pulses was sequentially activated firing every 8 µs for "on" cycles ranging from 20 µs to 250 µs, followed by quiescent "off" periods of equal duration, with an average power ≤0.8 mW on the sample plane. Timing details are described in greater detail in the Supporting Information. The net duty cycle of the QCL in this mode was maintained at <2% overall in single wavelength mode and ~4-5% for multichannel operation.

The visible and mid-IR lasers were focused to the sample plane through a zinc selenide (ZnSe) lens (f = 25mm, Thorlabs), calibrated using a clear pass USAF test grid (Edmund Optics). Divergence of the QCL beam was adjusted to align the visible and infrared focal planes. Images were collected by sample scanning with a piezoelectric stage (MadCityLabs Nano-BIO300). A photomultiplier tube (PMT, Hamamatsu) was used to collect fluorescence signal from the sample in epi-configuration. A 532nm notch filter and a 550nm long pass filter were used to block the incident 532nm light in front of the PMT. Signal from the PMT was passed through a tunable electronic band-pass filter (Ithaco M4302) centered at the QCL modulation frequency. The frequency-filtered signal was then passed through the lock-in amplifier (Stanford Research SR810). The amplified signal was then digitized (Alazar ATS9462), with images generated using MATLAB software written in-house. For spectral acquisition, F-PTIR signals for each wavelength channel were normalized to the transmitted intensity in the absence of a sample measured using an IR-photodiode (VIGO Systems).

Prior to implementation in microscopy, F-PTIR and FTIR spectra for thin liquid films were compared in

**Figure 2**. As can be seen in the figure, the F-PTIR spectra were generally in good agreement with absorbance spectra of both DMF and DMSO within the wavelength range accessible by the QCL used in these studies. Deviations between F-PTIR and FTIR for the low-frequency 1227 cm$^{-1}$ shoulder in the DMF spectrum are attributed to low laser power at those frequencies, with corresponding increases in F-PTIR measurement uncertainty.

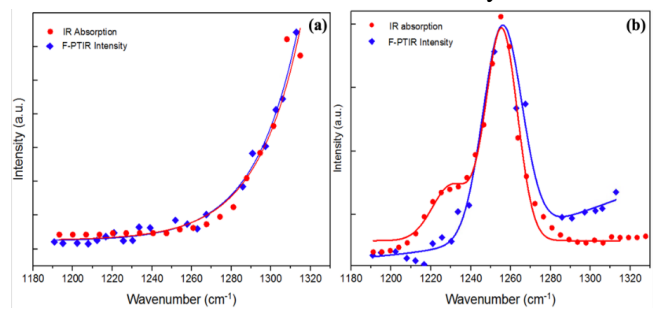

**Figure 2.** Comparison between F-PTIR intensity, normalized to QCL channel intensity, and mid-IR absorption spectrum for DMSO (a) and DMF (b) by FTIR.

Illustrated in **Figure 3(a)**, microscopy measurements were next performed to spectroscopically discriminate between particles in a mixture of two powdered materials: PEG and hydrated silica gel. From initial measurements of the unprocessed F-PTIR spectra of each of the two isolated materials, a compressive sensing binary wavelength mask was calculated to optimally discriminate between the two particles (detailed in the Supporting Information).[26] Following this calculation, F-PTIR microscopy measurements were performed serially using two mask patterns optimized to resolve PEG and silica gel particles. Consistent with these mask designs, images of the concentration maps exhibit little cross-talk and clear identification of the composition of individual particles (**Figure 3, b-d**).

Building on the results of these proof-of-concept studies with known materials, F-PTIR images of amorphous solid dispersions of ritonavir in PVPVA were acquired to explore the possible generation of amorphous phase-separated ritonavir-rich domains. Amorphous solid dispersions are widely used to address low aqueous solubility limitations of many active pharmaceutical ingredients (APIs) by dispersing the API within a polymer with established dissolution properties. In typical use-cases, dissolution of the polymer results in release of the API according to kinetics dictated by the polymer rather than the solid-state form of the API. However, prior studies of numerous ASD formulations have reported a "dropping off a cliff" phenomenon, shown in **Figure 4A**, in which API-loadings higher than a critical threshold result in abrupt *reductions* in the rate or extent of API dissolution.[27-28]

We conjecture that this reduction may be a consequence of phase separation and the formation of API-rich domains of low aqueous solubility. Building on our prior work,[29] spin-case films of ritonavir in PVPVA with trace (0.1 %) Nile red added as a fluorescence reporter were exposed to high humidity for 1 hour to induce phase-separation, then measured F-PTIR to determine composition within the phase-separated domains. The bright field, fluorescence, and F-PTIR microscopy measurements were performed, shown in **Figure 4 (c – d)**. From the results of the F-PTIR measurements, phase-separated domains exhibit clear enrichment in ritonavir based on the mid-infrared spectroscopy. To further corroborate the F-PTIR measurements and to assess the possible perturbation by the fluorescent marker, complementary label-free measurements of ritonavir/PVPVA mixtures were recorded by two-photon excited ultraviolet fluorescence (TPE-UVF) images, shown in **Figure 4B**. TPE-UVF leverages the weak but nonzero autofluorescence of ritonavir as an image contrast agent.[30] Comparisons of **Figures 4B** and **4D** indicate clear structural similarities between the phase-separated domains corresponding to ritonavir as identified by F-PTIR spectroscopy with those producing native TPE-UVF in the absence of a fluorescent marker.

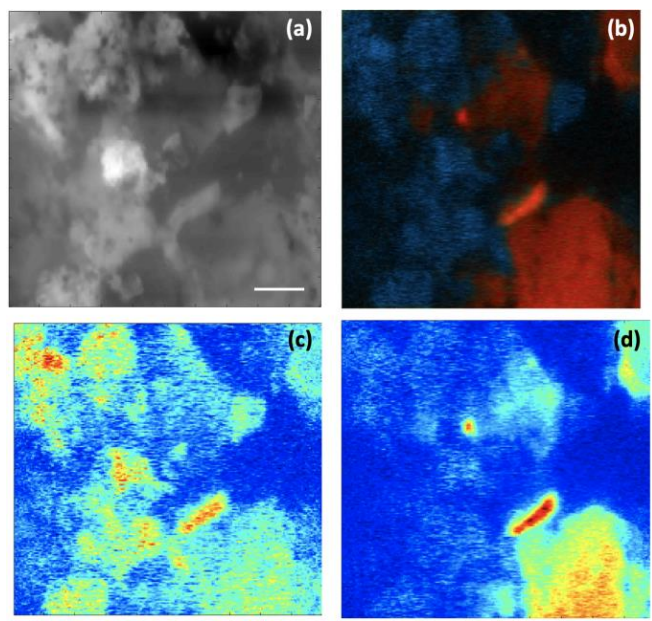

**Figure 3.** F-PTIR Microscopy of a mixture of silica gel (SG) and PEG particles. (a) – epi-fluorescence image of the field of view (FOV). (b) – components map within the FOV. Red regions correspond to species identified as SG; blue regions correspond to PEG. (c) and (d) – PEG and SG concentration maps after 10 NNMF iterations, respectively. Scale bar is 10 µm.

Spatial Fourier transform fluorescence recovery after photobleaching experiments (FT-FRAP) following Geiger *et al*.[31] confirm fluidity within the two phase-separated domains, with ~3-fold lower viscosity within the brightly fluorescent domains (see Supporting Information for details). Collectively, these results provide compelling evidence supporting amorphous/amorphous phase separation of API-rich domains within ritonavir /PVPVA films upon water uptake. These microscopic chemical microscopy measurements by F-PTIR further support an important role of liquid/liquid phase separation in dictating the kinetics of API dissolution from ASDs that determine bioavailability.

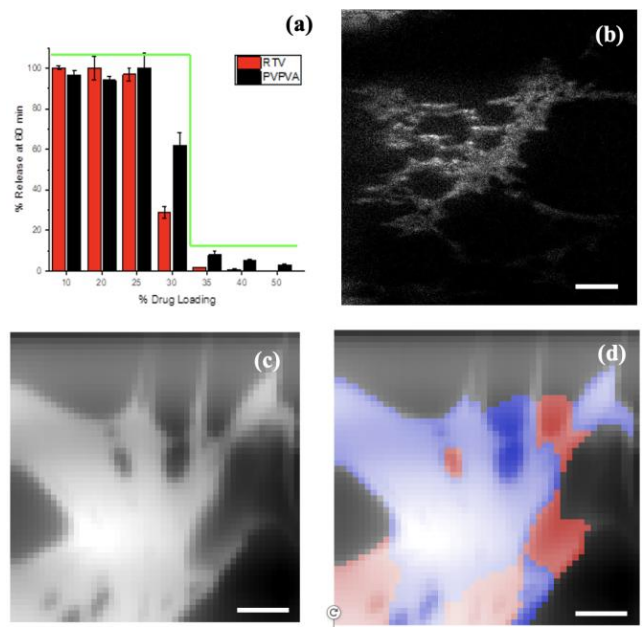

**Figure 4.** F-PTIR microscopy of amorphous phase separated ritonavir-PVPVA samples. (a) – An Illustration of the "dropping off the cliff" release phenomenon for RTV-PVPVA system[28]. (b) – Label-free TPE-UVF images of phase-separated film. (Scale bar is 100 μm). (c) – Epi-fluorescence image of the field-of-view used for F-PTIR measurements, with Nile Red partitioning to the bright domains. (d) - Classification of ritonavir-rich (blue) and PVPVA-rich (red) domains based on F-PTIR results. The image was segmented into 27 "superpixels" and each of them was assigned to either one of the components or background (see SI). (Scale bar is 15 μm).

The initial demonstration of F-PTIR herein lays a foundation to couple a diverse suite of fluorescence-based sensing strategies for signal transduction of mid-infrared absorption. Most directly, fluorescence provides specificity to labeled structures interrogated by IR absorption through F-PTIR; the fluorophore must reside within the thermal plume of the IR absorber in order to register a photothermal perturbation. The higher sensitivity of F-PTIR relative to O-PTIR suggests potential greater ease of integration into wide-field PTIR microscopy measurements. Furthermore, super-localization of fluorescence through STORM/PALM approaches can routinely yield spatial resolution on the order of 10's of nm in bulk media, suggesting potential promise for performing infrared spectroscopy by F-PTIR localized in 3D at comparable spatial resolution.

## ASSOCIATED CONTENT

**Supporting Information**.
The Supporting Information is available free of charge on the ACS Publications website, and includes the QCL specification and timing description; measurements of the co-propagating beam profile; F-PTIR microscope resolution determination; non-negative matrix factorization operation for optimizing the concentration map of silica gel and PEG; PVPVA – ritonavir phase-separated sample preparation; description of the super-pixel segmentation method for phase-separated sample; and FRAP analysis of viscosity differences between the phase-separated domains. (Supporting Infomation.docx)

## AUTHOR INFORMATION

**Corresponding Author**
gsimpson@purdue.edu.

**Author Contributions**
†These authors contributed equally.

## ACKNOWLEDGMENT

ML, AR, and GJS gratefully acknowledge funding from the National Science Foundation through an NSF-GOALI award (CHE-1710475). ML and AR also acknowledge support from the NSF Center for Bioanalytic Metrology (IIP-1916691). RY acknowledges the Agency for Science, Technology and Research (A*STAR), Singapore for funding. Pendar Technologies acknowledges support from US Army under SBIR topic A14A-T015.